\documentclass[10pt,superscriptaddress,twocolumn,showpacs,pra,longbibliography]{revtex4-1}
\usepackage{bm} 
\usepackage{graphicx} 
\usepackage{amsmath}
\usepackage{amsthm}
\usepackage{amssymb} 
\usepackage{comment} 
\usepackage{epstopdf} 
\usepackage{hyperref}
\usepackage{hypernat}
\usepackage{amsfonts}
\usepackage{color} 
\usepackage{subfigure}
\usepackage[english]{babel}
\usepackage{enumitem}
 \usepackage[bb=boondox]{mathalfa}

\newcommand{\bra}[1]{\langle #1 |}
\newcommand{\ket}[1]{| #1 \rangle}

\newcommand {\be}{\begin{equation}}
\newcommand {\ee}{\end{equation}}

\newcommand\s{{\text{stoq}}}
\newcommand\spf{{\text{VGP}}}
\newcommand\abs{{\text{ABS}}}

\newcommand{\ba}{\begin{eqnarray}}
\newcommand{\ea}{\end{eqnarray}}
\newcommand\tr{{\mbox{Tr\,}}}
\newcommand{\ignore}[1]{}

\newcommand{\bes} {\begin{subequations}}

\newcommand{\ees} {\end{subequations}}

\newcommand\sgn{{\text{sgn}}}

\DeclareRobustCommand\openzero{\leavevmode{0\kern-.55em0}}
\mathchardef\minus="002D

\usepackage[margin=2cm,nofoot]{geometry}

\newcommand{\beq}{\begin{eqnarray}}
\newcommand{\eeq}{\end{eqnarray}}

\newcommand{\e}{{e}}

\renewcommand{\Re}{\operatorname{Re}}

\newcommand {\bea}{\begin{eqnarray}}
\newcommand {\eea}{\end{eqnarray}}

\begin{document}

\title{Determining QMC simulability with geometric phases}
\author{Itay Hen}
\affiliation{Department of Physics and Astronomy and Center for Quantum Information Science \& Technology, University of Southern California, Los Angeles, California 90089, USA}
\affiliation{Information Sciences Institute, University of Southern California, Marina del Rey, California 90292, USA}
\email{itayhen@isi.edu}
\begin{abstract}
\noindent Although stoquastic Hamiltonians are known to be simulable via sign-problem-free quantum Monte Carlo (QMC) techniques, the non-stoquasticity of a Hamiltonian does not necessarily imply the existence of a QMC sign problem. We give a sufficient and necessary condition for the QMC-simulability of Hamiltonians in a fixed basis in terms of geometric phases associated with the chordless cycles of the weighted graphs whose adjacency matrices are the Hamiltonians. We use our findings to provide a construction for non-stoquastic, yet sign-problem-free  and hence QMC-simulable, quantum many-body models. We also demonstrate why the simulation of truly sign-problematic models using the QMC weights of the stoquasticized Hamiltonian is generally sub-optimal. We offer a superior alternative.  
\end{abstract}

\maketitle

\section{Introduction}
The concept of stoquasticity, first introduced by Bravyi et al.~\cite{Bravyi:QIC08}, is a key definition in both quantum Monte Carlo simulations and computational complexity theory. A Hamiltonian is dubbed stoquastic with respect to a given basis if and only if all its off-diagonal elements in that basis are nonpositive. Otherwise it is referred to as non-stoquastic. In complexity theory, the complexity class StoqMA, associated with the problem of deciding whether the ground-state energy of stoquastic local Hamiltonians is above or below certain values, is expected to be strictly contained in the complexity class QMA, which deals with general local Hamiltonians~\cite{2002quant.ph.10077A,TCS-066,Bravyi:QIC08}. StoqMA is also an essential part of the complexity classification of local Hamiltonian problems~\cite{doi:10.1137/140998287}.

In the field of quantum Monte Carlo (QMC) simulations~\cite{Landau:2005:GMC:1051461,newman}, the partition function of stoquastic Hamiltonians can always be written as a sum of efficiently computable strictly positive weights~\cite{DBLP:journals/qic/Bravyi15,elucidating,pmr}. As a consequence, such Hamiltonians do not suffer from a `sign problem'~\cite{Wiese-PRL-05,signProbSandvik}, i.e., from the existence of negative summands, which greatly impede the convergence of QMC algorithms~\cite{Wiese-PRL-05,marvianLidarHen,klassenMarvian,signProbSandvik}. 

That stoquasticity leads to sign-problem-free (SPF) representations of quantum many-body physical models has served as the main motivation in numerous recent studies that examine the conditions under which various different classes of non-stoquastic Hamiltonians can be unitarily transformed to equivalent stoquastic~\cite{marvianLidarHen,Klassen2019twolocalqubit,klassenMarvian,2020arXiv200711964I}, or `minimally non-stoquastic'~\cite{eisertEasing} representations.

Although the dichotomy between stoquastic and non-stoquastic Hamiltonians is very often convenient, stoquasticity is 
not (nor was it intended to be) the property that differentiates QMC-simulable (i.e., SPF) Hamiltonians from non-simulable ones. Although the stoquasticity of a Hamiltonian implies an SPF decomposition of the partition function, the converse, namely that non-stoquasticity leads to a sign-problematic representation, is not necessarily true~\cite{elucidating}. 

A natural question thus arises: What property of a Hamiltonian makes it QMC-simulable? In this study, we answer this question in terms of the geometric phases associated 
with the chordless cycles of a weighted graph whose adjacency matrix is the Hamiltonian in question. 
Our result also has practical significance. First and foremost, it clearly illustrates that `curing' the sign problem of a model, i.e., finding a unitary transformation that produces an SPF representation for it, is markedly different from curing non-stoquasticity (finding unitary transformations that make the Hamiltonian stoquastic). In fact, our result shows that the latter approach, which is the current standard practice~\cite{marvianLidarHen,Klassen2019twolocalqubit,klassenMarvian,2020arXiv200711964I,eisertEasing}, should not be used toward rendering a Hamiltonian simulable.  In addition, our result demonstrates 
that `stoquastization' of sign-problematic Hamiltonians, the method normally used for assigning positive weights to QMC configurations is, in general, a sub-optimal choice; a superior alternative can be given in terms of the geometric phases of the Hamiltonian.
We further demonstrate the above by introducing a class of quantum many-body models that, while being non-stoquastic, are SPF and perfectly simulable by QMC techniques.

The paper is organized as follows. In Sec.~\ref{sec:sp} we present a generic partition function decomposition focusing on the signs of its summands and their origins, which we trace back to geometric phases associated with the graph structure of the Hamiltonian. Based on the observations made in the preceding section, we provide in Sec.~\ref{sec:nspf} a construction for non-stoquastic sign-problem-free Hamiltonians. In Sec.~\ref{sec:stoq} we discuss QMC simulations of sign-problematic Hamiltonians. Conclusions are given in Sec.~\ref{sec:conc}.


\section{Emergence of the sign problem as a nontrivial geometric phase\label{sec:sp}}

We now examine in detail the origins of the sign problem in QMC simulations. By doing so, we are able to differentiate 
QMC-simulable Hamiltonians (in a given basis) from non-simulable ones.

\subsection{Permutation matrix representation of Hamiltonians}

We start our derivation by considering a Hamiltonian $H$ given in a basis $\mathcal{B}=\{ |z\rangle \}$, which we refer to as the computational basis, and write the Hamiltonian in `permutation matrix representation' (PMR)~\cite{ODE,ODE2,pmr}, i.e.,
as a sum
\beq \label{eq:basic}
H=\sum_{j=0}^M \tilde{P}_{j} =\sum_{j=0}^M D_j P_j  \,,
\eeq
where $\{ \tilde{P}_j\}$ is a set of $M+1$ distinct generalized permutation matrices~\cite{gpm}, i.e., matrices with precisely one nonzero element in each row and each column (this condition can be relaxed to allow for zero rows and columns). 
Each operator $\tilde{P}_j$ can be written, without loss of generality, as $\tilde{P}_j=D_j P_j$ where $D_j$ is a diagonal matrix
and $P_j$ is a  permutation matrix with no fixed points (equivalently, no nonzero diagonal elements) except for the identity matrix $P_0=\mathbb{1}$. We will call the diagonal matrix $D_0$ the `classical Hamiltonian'.
Each term $D_j P_j $ obeys 
$D_j P_j | z \rangle = d_{z'}^{j} | z' \rangle$ where $d_{z'}^{j}$ is a possibly complex-valued coefficient and $|z'\rangle \neq |z\rangle$ is a basis state.

\subsection{Partition function decomposition}

\begin{figure}[htp]
\includegraphics[width=.48\textwidth]{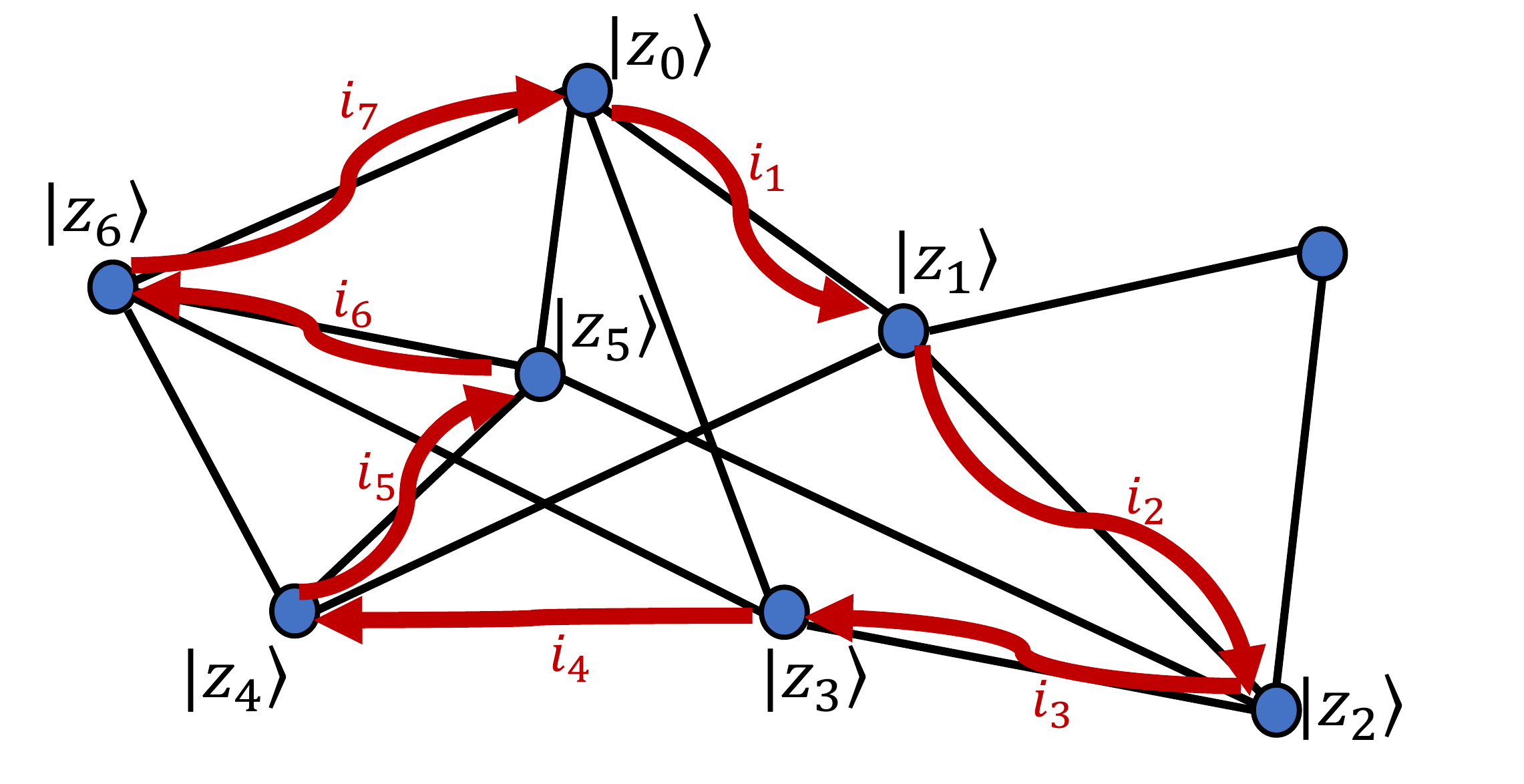}
\caption{\label{fig:hyper} Every summand $W_{(z,{\bf{i}}_q)}$ in the partition function decomposition is associated with a closed walk on the hypercube of basis states (a subgraph of which is shown here for illustration). The walk is determined by the action of the permutation operators of the configuration, represented by the sequence $S_{{\bf{i}}_q}=P_{i_q} \cdots P_{i_2} P_{i_1}$, on the initial basis state $|z_0\rangle$. Every node $|z_j\rangle$ has an associated classical energy $E_{z_j}$. 
}
\end{figure}

Having cast $H$ in PMR form, we next derive an expression for the partition function $Z=\tr\left[ \e^{-\beta H} \right]$.
Expanding the exponential in a Taylor series in the inverse-temperature $\beta$, $Z$ can be written as a triple sum over all basis states $|z\rangle$, the expansion order $q$ which ranges from 0 to infinity and the (unevaluated) products \hbox{$S_{{\bf{i}}_q} = P_{i_q} \ldots P_{i_2} P_{i_1}$} of $q$ off-diagonal operators. Here we use the multiple index ${\bf i}_q = (i_1,\ldots,i_q)$ where each individual index $i_j$ ranges from $1$ to $M$. In this notation, the empty sequence $S_{\bf{i}_0}$ corresponds to the identity operation.
After some algebra (the reader is referred to Ref.~\cite{pmr} for a full derivation), the partition function attains the form
 \beq \label{eq:z1}
Z =\sum_{\{z\}} \sum_{q=0}^{\infty}  \sum_{{\bf{i}}_q}  D_{(z,{\bf{i}}_q)} \bra{z} S_{{\bf{i}}_q} \ket{z}   \e^{-\beta [E_{z_0},\ldots,E_{z_q}]} \,,\nonumber\\
\eeq 
where $\{S_{{\bf{i}}_q}\}$ is the set of all (unevaluated) products \hbox{$P_{i_q} \ldots P_{i_2} P_{i_1}$} of size $q$  and $e^{-\beta[E_{z_0},\ldots,E_{z_q}]}$ is a divided difference of $f(\cdot)=\e^{-\beta (\cdot)}$ (see Appendix~\ref{app:divDiff} for an overview) with inputs $[E_{z_0},\ldots E_{z_q}]$~\cite{dd:67,deboor:05}. 
The energies $E_{z_i}=\langle z_i | D_0|z_i\rangle$ (where $i=0,\ldots, q$) are the `classical' energies of the states $|z_0\rangle, \ldots, |z_q\rangle$, which are in turn obtained from the action of the ordered $P_j$ operators in the sequence $S_{{\bf{i}}_q}$ on $|z_0\rangle$, then on $|z_1\rangle$, and so forth.
Explicitly, $|z_0\rangle=|z\rangle, P_{i_1}|z_0\rangle=|z_1\rangle, P_{i_2}|z_1\rangle=|z_2\rangle$, etc. The sequence of basis states $\{|z_i\rangle \}$ may be viewed as a closed walk~\cite{gst} of length $q$ on the hypercube of basis states. See Fig.~\ref{fig:hyper} for an illustration. The expression $\bra{z} S_{{\bf{i}}_q} \ket{z}$ is $1$ if  $ S_{{\bf{i}}_q}$ evaluates to the identity operator. Otherwise it is zero and can be removed from the sum. 
(Note that \hbox{$|z_j\rangle=P_{i_j} \ldots P_{i_2} P_{i_1}|z\rangle$} in principle should be denoted $|z_{ (i_1 , \ldots ,i_j ) } \rangle$. We use a simplified notation so as not to overburden the notation.) 
Additionally, we denote
\beq
D_{(z,{\bf{i}}_q)}=\prod_{j=1}^q d^{({\bf{i}}_j)}_{z_j}\,,
\eeq
where $d^{({\bf{i}}_j)}_{z_j} = \langle z_j | D_{i_j}|z_j\rangle$ are off-diagonal elements of $H$.

\begin{figure}[htp]
\includegraphics[width=.48\textwidth]{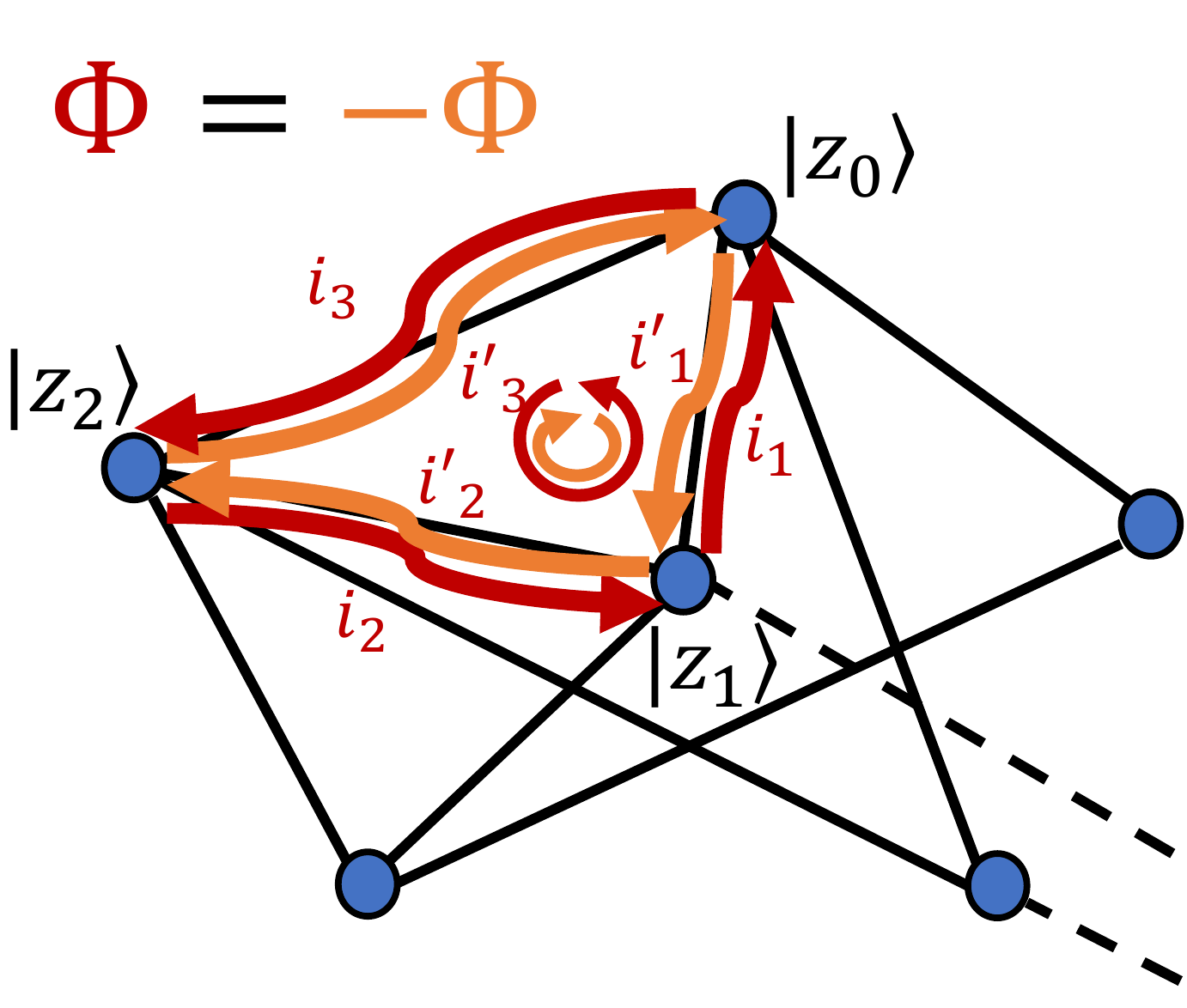}
\caption{\label{fig:pathWalksConj} Every closed walk has a conjugate closed walk traversing the same path but in the reverse direction. The weight of a walk is the complex conjugate of the weight of its conjugate walk. This observation allows us to disregard the imaginary parts of these partition function summands that cancel out. The permutation operators that generate a closed walk are the inverses of the permutation operators that generate its conjugate and appear in reverse order.}
\end{figure}

The summands of the partition function decomposition are thus 
\hbox{$D_{(z,{\bf{i}}_q)}  \e^{-\beta [E_{z_0},\ldots,E_{z_q}]}$}.
These can, in general, be complex-valued, despite the partition function being real (the $d^{({\bf{i}}_j)}_{z_j}$ can, in the general case, be complex).  However, we note that for every configuration $(z,{\bf{i}}_q)$ there is a conjugate configuration $(z,{\bf{i}}'_q)$,  which produces the conjugate weight $W_{(z,{\bf{i}}'_q)}=\overline{W}_{(z,{\bf{i}}_q)}$. Explicitly, for every closed walk $S_{{\bf{i}}_q} =P_{i_q} \ldots  P_{i_2} P_{i_1}$ there is a conjugate walk in the reverse direction, whose operator sequence is $S^\dagger_{{\bf i}_q} = P_{i_1}^{-1} P_{i_2}^{-1} \ldots P_{i_q}^{-1}$ (see Fig.~\ref{fig:pathWalksConj} for an illustration).  The imaginary parts of the complex-valued summands therefore do not contribute to the partition function and may be disregarded altogether. We may therefore take
\beq\label{eq:W}
W_{(z,{\bf{i}}_q)} = \Re[D_{(z,{\bf{i}}_q)}]  \e^{-\beta [E_{z_0},\ldots,E_{z_q}]} \,
\eeq
as the summands of the expansion. 

We can now examine the condition for the positivity of the summands $W_{(z,{\bf{i}}_q)}$. First, we note that the term $e^{-\beta[E_{z_0},\ldots,E_{z_q}]}$ is positive (negative) for even (odd) values of $q$, the length of the walk (see Appendix~\ref{app:posDivDiff} for a short proof) and so the sign of a summand can be simplified to
\beq \label{eq:PosCond}
\sgn \left[W_{(z,{\bf{i}}_q)}\right]=\sgn \Re[\prod_{j=1}^q (-d^{({\bf{i}}_j)}_{z_j})]
\eeq
i.e., the partition function expansion will admit a negative weight if and only if there exists a closed walk on the hypercube of basis states along which $\Re\left[\prod_{j=1}^q (-d^{({\bf{i}}_j)}_{z_j})\right]<0$. 

\subsection{From walks to cycles}

The expression derived above for the sign of a partition function summand can be given a more geometrical meaning if we write the off-diagonal matrix elements  $d^{({\bf{i}}_j)}_{z_j}$ in polar coordinates, i.e.,
\hbox{$d^{({\bf{i}}_j)}_{z_j}=-r^{({\bf{i}}_j)}_{z_j} \e^{-i \phi^{({\bf{i}}_j)}_{z_j}}$} (note the extra minus sign introduced for notational convenience). 
This allows us to write the sign of a summand as 
\bea
\sgn \left[W_{(z,{\bf{i}}_q)}\right] &=& \sgn \Re [\prod_{j=1}^q (\e^{-i \phi^{({\bf{i}}_j)}_{z_j}})]=\sgn \cos \Phi_{(z,{\bf{i}}_q)}\,,\nonumber\\
\eea
where we define
\beq\label{eq:phase}
\Phi_{(z,{\bf{i}}_q)} =\sum_{j=1}^q \phi^{({\bf{i}}_j)}_{z_j}
\eeq
 as the overall \emph{geometric phase} associated with the summand walk. 
For a Hamiltonian to be SPF, one must ensure that all closed walks have a positive $\cos \Phi_{(z,{\bf{i}}_q)}$. 

\begin{figure}[htp]
\includegraphics[width=.48\textwidth]{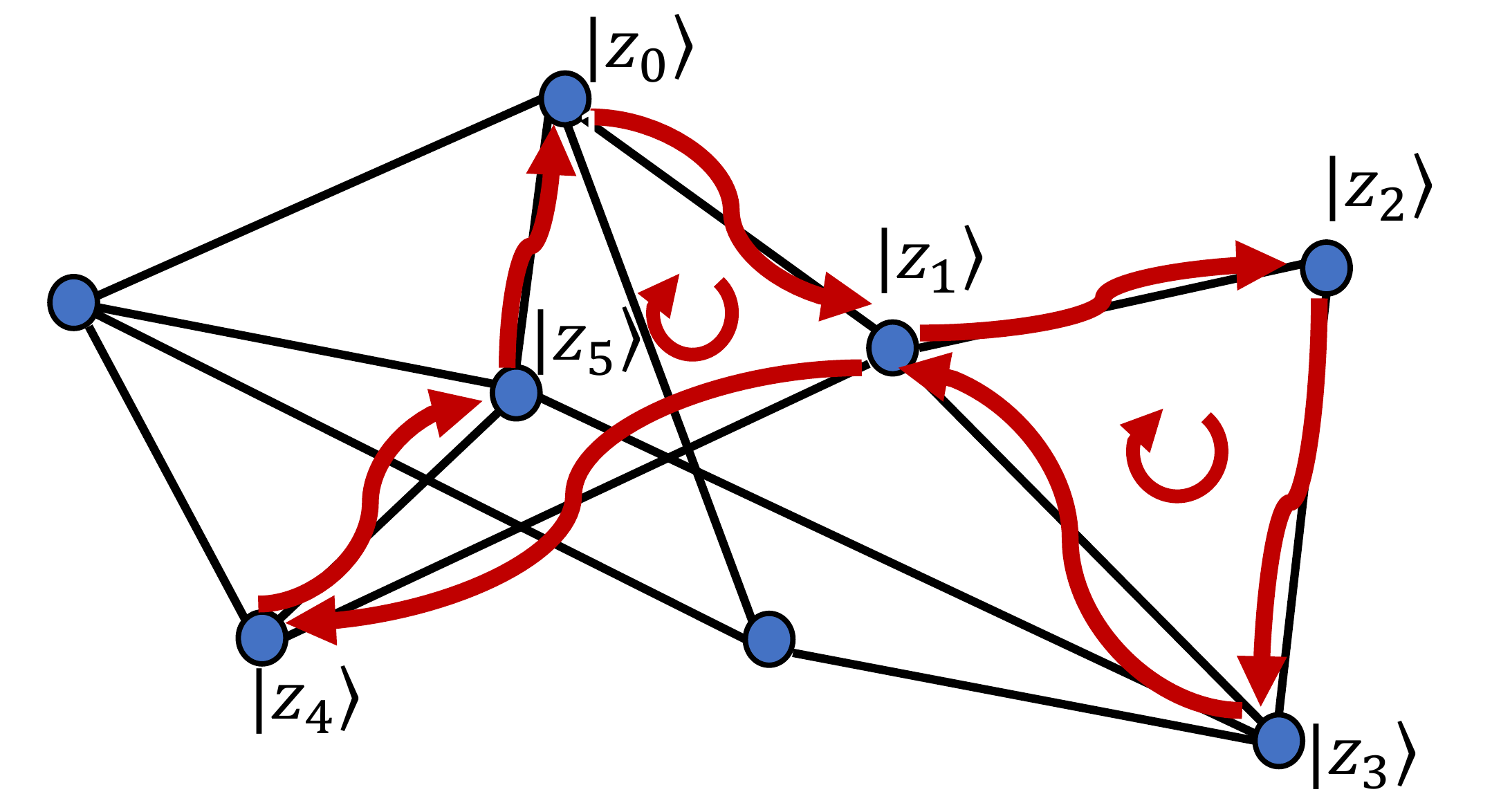}
\caption{\label{fig:pathWalks} Composite walks can be viewed as a concatenation of cycles (cyclic paths on a graph). Above, the closed walk $z_0 \to z_1 \to \cdots \to z_5 \to z_1$ can be decomposed to two cycles: $z_1 \to z_2 \to z_3 \to z_1$ and $z_0 \to z_1 \to z_4 \to z_5 \to z_0$. The overall phase associated with a closed walk is the sum of the phases associated with its constituent cycles.}
\end{figure}

The number of closed walks on the hypercube of basis states for any given Hamiltonian is infinite, even for finite graphs, since nodes can be revisited. Nonetheless, every closed walk may be viewed as the concatenation of distinct closed paths or \emph{cycles} (equivalently, non-repeating walks).  Moreover, the phase of any given closed walk may be written as the sum of the phases of its constituent cycles (see Fig.~\ref{fig:pathWalks} for an illustration).

\begin{figure}[htp]
\includegraphics[width=.48\textwidth]{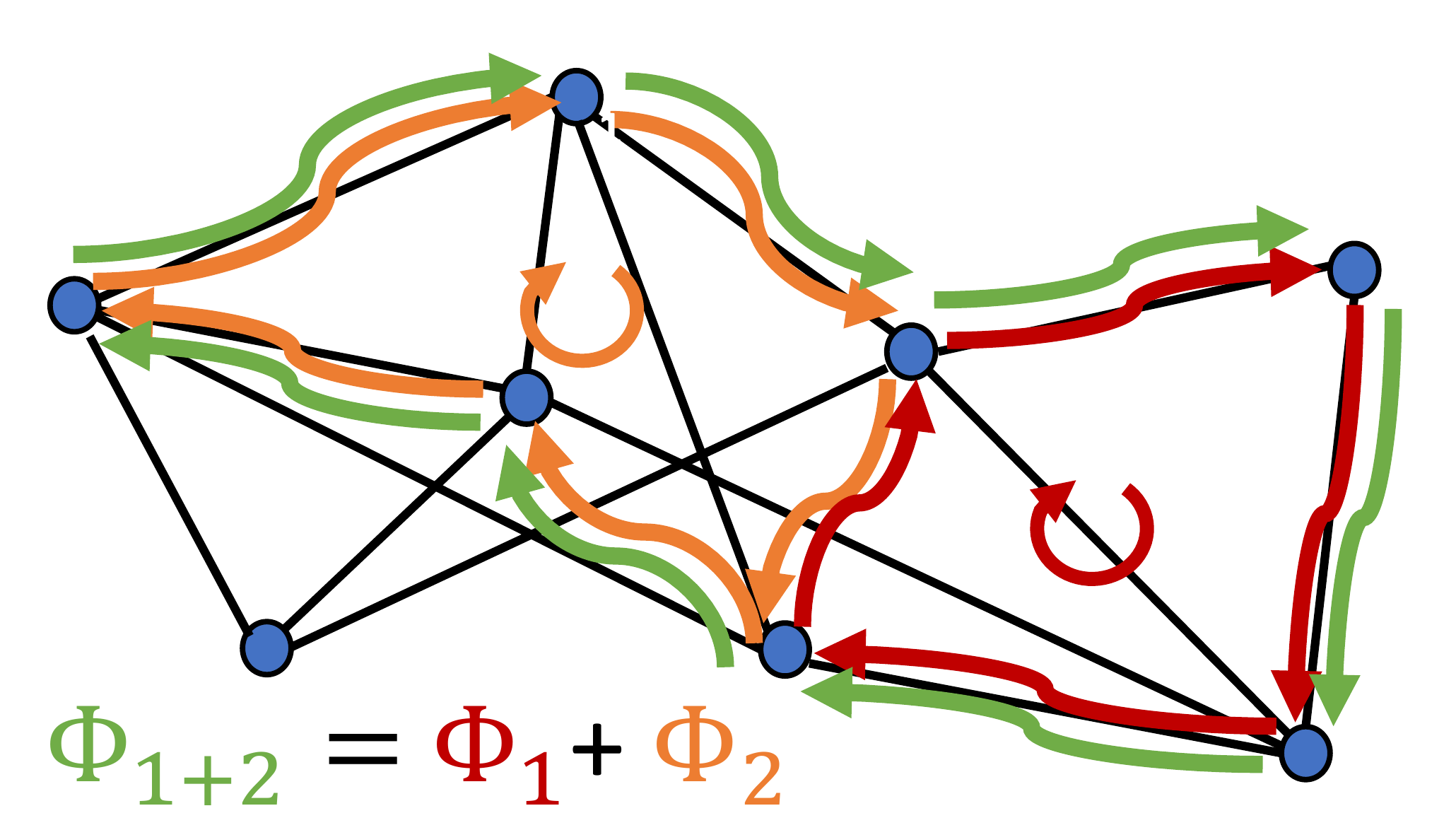}
\caption{\label{fig:pathWalks2}  The geometric phase associated with a composite cycle (green) is the total sum of the phases of its sub-cycles (red and orange).  }
\end{figure}

Furthermore, long cycles may be viewed as the concatenation of smaller, basic or `induced' (also referred to as `chordless') cycles~\cite{graphTheoryBook} that cannot in turn be decomposed to yet shorter cycles (see Fig.~\ref{fig:pathWalks2}) and the overall phase associated with long cycles is the sum total of the phases of its constituent induced cycles.

\subsection{Geometric phases of the chordless cycles of SPF Hamiltonians}

Next, we answer the question of what values closed-walk phases can take in order to ensure that a given Hamiltonian has no sign problem. Let us consider an induced cycle with a total phase $\Phi$. As per Eq.~(\ref{eq:phase}), as long as $\cos\Phi \geq 0$, the weight associated with the cycle is positive [i.e., $\Phi \in (-\pi/2,\pi/2)$ modulo $2\pi$]. However, since walks consisting of repeated concatenations of the original cycle are also legitimate walks, one must therefore also ensure that  $\cos \Phi_m=\cos m \Phi \geq 0$ for every natural number $m$. For this much stronger condition to hold, the induced-cycle phase must be $\Phi=0$ modulo $2\pi$ (see Appendix~\ref{app:cosmx}), i.e., the cycle must have a vanishing geometric phase (VGP). 


Having observed  that the chordless cycles of a Hamiltonian must have vanishing geometric phases for their associated weights to be positive (we call these VGP Hamiltonians), it is straightforward to generalize the condition to include longer cycles and, in fact, all closed walks: If all cycles of a Hamiltonian have vanishing phases then all closed walks, which are concatenations of cycles, have vanishing phases as well. We thus conclude that \emph{for general Hamiltonians to have no sign problem in a given basis, the geometric phases of the chordless cycles of their Hamiltonian must be zero modulo $2 \pi$}. This condition is necessary and sufficient. 


In the \emph{special} case where all $d^{({\bf{i}}_j)}_{z_j}$, which correspond to off-diagonal Hamiltonian elements, are negative and hence have zero phase $\phi^{({\bf{i}}_j)}_{z_j}=0$, we obtain $\Phi_{(z,{\bf{i}}_q)}=0$ for every closed walk. This in turn implies  that the weights associated with all walks are positive and the Hamiltonian will be SPF. This is the case of stoquastic Hamiltonians. 

To illustrate the difference between stoquasticity and VGP, in the next section we introduce a class of Hamiltonians that are non-stoquastic yet VGP, making them QMC-simulable sign-problem-free Hamiltonians. 

\section{Non-stoquastic sign-problem-free Hamiltonians\label{sec:nspf}}

Let us consider a general $N$-by-$N$ stoquastic Hamiltonian $H^{(\s)}$, i.e., a general symmetric matrix with nonpositive off-diagonal elements (explicitly, $H^{(\s)}_{ij} \leq 0$ for all $i,j \in \{1,\ldots,N\}$). 
We next define
\beq\label{eq:spfH}
H^{(\spf)}= \e^{i \Theta} H^{(\s)} \e^{-i \Theta} \,,
\eeq
where $\e^{i \Theta} = diag\{\theta_1,\ldots,\theta_N\}$ is a diagonal matrix of phases.  
The off-diagonal elements of $H^{(\spf)}$ are
\beq
H^{(\spf)}_{ij} = H^{(\s)}_{ij}  \e^{i (\theta_i -\theta_j)} \,,
\eeq
and so $H^{(\spf)}$ will generally be non-stoquastic. Nonetheless, it is straightforward to show that $H^{(\spf)}$ above is VGP. This is because for any closed walk, i.e., a sequence of off-diagonal elements, we have 
\bea
&&(-H^{(\spf)}_{ij}) (-H^{(\spf)}_{jk}) \cdots (-H^{(\spf)}_{ml})(-H^{(\spf)}_{li})\nonumber\\
&=& 
(-H^{(\s)}_{ij} \e^{i (\theta_i -\theta_j)} ) (-H^{(\s)}_{jk}\e^{i (\theta_j-\theta_k)} ) \cdots
\\\nonumber
&=&(-H^{(\s)}_{ij}) (-H^{(\s)}_{jk}) \cdots (-H^{(\s)}_{ml} )(-H^{(\s)}_{li}) \geq 0\,.
\eea
The converse is also true: If a Hamiltonian is VGP, it can always be unitarily transformed to stoquastic form via a rotation by a phase matrix. To show this, consider a VGP Hamiltonian whose off-diagonal elements are written in polar form \hbox{$H^{(\spf)}_{ij} = -r_{ij} \e^{i \phi_{ij}}$}. Its phases obey $\phi_{ji}=-\phi_{ij}$ (hermiticity) and \hbox{$(\phi_{ij} + \phi_{jk}+ \cdots +\phi_{ml} + \phi_{li}) =0$} modulo $2 \pi$ (VGP) provided that all of $r_{ij}, r_{jk}, \ldots, r_{ml}, r_{li}>0$.  
We will prove that there is always a rotation matrix $\e^{-i \Theta}$ that can rotate $H^{(\spf)}$ to stoquastic form; that is, we will show that there is always a choice $\{\theta_1,\ldots,\theta_N\}$ such that $\phi_{ij}+\theta_i-\theta_j=0$ modulo $2 \pi$ for all $i,j$. To do that, we will prove that are at most $N-1$ independent phases, which if `cured' imply the curing of all other phases, owing to the VGP condition. Since there are $N-1$ independent independent choices for $\theta_i - \theta_j$, we can choose  $\theta_i - \theta_j = -\phi_{ij}$ for all independent phases, thereby curing all phases. 

To show that there are at most $N-1$ independent phases, let us assume for simplicity that all off-diagonal elements are nonzero, that is $r_{ij}>0$ for all $i \neq j$ (the proof is similar even in the absence of this condition). It is easy to show that the $N-1$ phases $\phi_{12},\ldots,\phi_{1N}$ determine all other phases. From hermitically, we have $\phi_{m1}=-\phi_{1m}$ for all $m=2,\ldots ,N$. Next, consider the phases of the cycles $(-H^{(\spf)}_{1n}) (-H^{(\spf)}_{nm}) \cdots (-H^{(\spf)}_{m1})$. VGP imposes $\phi_{1n} + \phi_{nm} + \phi_{m1} = 0$ for $n,m \neq 1$, which in turn implies that all $\phi_{nm}$ are defined modulo $2\pi$. If all $\phi_{1m}$ are cured after rotation then so will all $\phi_{m1}$, and due to VGP, the same is true for all $\phi_{nm}$. 

Note that the curing transformation $\e^{-i \Theta}$ is a nonlocal one in general, and finding it  requires solving a linear set of (up to) $N-1$ equations. As previously noted however, there is no actual need for `curing' non-stoquastic VGP Hamiltonians because they are sign-problem-free to begin with. 

\section{QMC simulations without stoquastization\label{sec:stoq}}

The VGP Hamiltonians discussed in the previous section illustrate an additional point of practical significance to the QMC simulation of (truly) sign-problematic systems that admit negative terms in the partition function decomposition.
For those, it has become common practice to choose as QMC weights the positive terms of the `stoquasticized' version of the Hamiltonian (see, e.g., Refs.~\cite{Wiese-PRL-05,Crosson2020designing}). Explicitly, the weights of $H$ are taken to be those of $H^{(\s)}$, the stoquasticized version of the Hamiltonian obtained by $H_{ij}^{(\s)} = - |H_{ij}|$, which are guaranteed to be positive (this choice is sometimes referred to as `bosonization'). In our notation, the positive stoquasticized weights are given by
\beq
W_{(z,{\bf{i}}_q)}^{(\s)} =\left| D_{(z,{\bf{i}}_q)}  \e^{-\beta [E_{z_0},\ldots,E_{z_q}]}\right| \,.
\eeq
The above choice is, however, in general not optimal. To understand why this is so, we examine the `weighted sign', a measure of how adverse the sign problem is in a QMC simulation (QMC convergence times are inversely proportional to the value of the weighted sign). This quantity, which we denote as $\langle \text{sgn}\rangle$, is simply the ratio of the sum of `true' weights $W_{(z,{\bf{i}}_q)}$ as per Eq.~(\ref{eq:W}) (i.e., the partition function of the system in question) to the sum of the weights chosen for the simulation~\cite{Wiese-PRL-05,elucidating}. 

If weights are chosen via stoquastization, the weighted sign becomes
\beq\label{eq:sgnDef}
\langle \text{sgn}\rangle_{\s}=\frac{\sum_{(z,{\bf{i}}_q)} W_{(z,{\bf{i}}_q)}}{\sum_{(z,{\bf{i}}_q)} W_{(z,{\bf{i}}_q)}^{(\s)}}\,.
\eeq 
However, a more appropriate choice of positive weights is to place the absolute value on the cosine of the total phase of each walk, namely, 
\beq\label{eq:phase}
W_{(z,{\bf{i}}_q)}^{(\abs)}=|W_{(z,{\bf{i}}_q)}|=|\cos \Phi_{(z,{\bf{i}}_q)}| W_{(z,{\bf{i}}_q)}^{(\s)} \,.
\eeq
The above choice leads to a weighted sign of  
\beq
\langle \text{sgn}\rangle_{\abs}
=\frac{\sum_{(z,{\bf{i}}_q)} W_{(z,{\bf{i}}_q)}}{\sum_{(z,{\bf{i}}_q)} W_{(z,{\bf{i}}_q)}^{(\abs)}}=
\frac{\sum_{(z,{\bf{i}}_q)} W_{(z,{\bf{i}}_q)}}{\sum_{(z,{\bf{i}}_q)}  |\cos \Phi_{(z,{\bf{i}}_q)}| W_{(z,{\bf{i}}_q)}^{(\s)}}\nonumber\\
\eeq
from which it is clear that 
\beq
\langle \text{sgn}\rangle_\abs \geq \langle \text{sgn}\rangle_{\s}\,,
\eeq 
with equality only when 
$\langle \text{sgn}\rangle_\abs=1$, i.e., for SPF Hamiltonians. 
In all other cases, where \hbox{$\langle \text{sgn}\rangle_{\abs}<1$}, the inequality is a strict one and the sign problem becomes provably less severe for the latter alternative. 
In fact, in the low-temperature limit we expect \hbox{$\lim_{\beta \to \infty} \langle \text{sgn}\rangle_\abs/\langle \text{sgn}\rangle_\s =\infty$} as both quantities decay exponentially fast to zero as a function of $\beta$~\cite{Wiese-PRL-05,elucidating,eisertEasing}, but at different rates. 

\section{Summary and conclusions\label{sec:conc}}

We gave a necessary and sufficient condition for the QMC-simulability of Hamiltonians in a given basis. 
We found that if and only if all the geometric phases of the chordless cycles of the weighted graph whose adjacency matrix is the Hamiltonian are zero (modulo $2\pi$), the simulation will be sign-problem-free. To further distinguish simulability from non-stoquasticity, we presented a construction for non-stoquastic yet sign-problem-free Hamiltonians. 

We also showed that simulating sign-problematic Hamiltonians by choosing the weights to be the summands of the analogous `stoquasticized' Hamiltonian, i.e., the stoquastic analogue of the Hamiltonian, is a sub-optimal choice in general. We provided a more suitable choice for said weights in terms of the absolute values of the cosines of the geometric phases of the simulated Hamiltonian, the weighted sign due to which is bounded from below by the weighted sign generated by the stoquastic choice. We can therefore expect that the choice of positive weights proposed here will allow for more efficient QMC simulations of sign-problematic quantum many-body models and, in turn,  will enable the study of considerably larger systems than possible today by stoquastization. 

Our study advocates the studying of the conditions under which Hamiltonians can be unitarily transformed to VGP form, rather than to the less relevant stoquastic form, which is the common practice today~\cite{marvianLidarHen,Klassen2019twolocalqubit,klassenMarvian,2020arXiv200711964I,eisertEasing} for curing the sign problem.
Especially worth mentioning in this context are recent results that have shown that for certain classes of Hamiltonians, `curing' non-stoquasticity, i.e., finding unitary transformations that transform the Hamiltonian to stoquastic form, is an intractable (NP-hard) task. The fact that non-stoquasticity does not imply non-simulability suggests that demanding the 
curing of non-stoquasticity is in general too excessive and that the weaker condition of seeking transformations to VGP form are more appropriate. 

Developing a true understanding of the nature of the QMC sign problem will have implications across all branches of the physical sciences and is crucial to the potential resolution of the problem. We hope that this study will provide a useful framework for making progress in this context. 

Other directions of research that we believe are worth pursuing are studying the extent to which the concept of vanishing geometric phases is also relevant for results pertaining to the use of stoquasticity in complexity theory~\cite{Bravyi:QIC08,2002quant.ph.10077A,TCS-066,doi:10.1137/140998287} and in other domains specifically in determining the positivity of Hamiltonian ground states~\cite{perron} and the bounding of their spectral gaps~\cite{2018arXiv180406857J}. In Ref.~\cite{2018arXiv180406857J} for example, a similar generalization of stoquasticity was proposed in a different context. The question of whether all results pertaining to stoquasticity hold for VGP as well is an interesting one.  Another question of interest is whether the complexity of verifying that a given Hamiltonian is VGP is different than the complexity associated with verifying that it is stoquastic. We leave that for future work. 


\begin{acknowledgements}
We thank Elizabeth Crosson, Michael Jarret and Milad Marvian for valuable comments and discussions. 
The research is based upon work (partially) supported by the Office of
the Director of National Intelligence (ODNI), Intelligence Advanced
Research Projects Activity (IARPA) and the Defense Advanced Research Projects Agency (DARPA), via the U.S. Army Research Office
contract W911NF-17-C-0050. The views and conclusions contained herein are
those of the authors and should not be interpreted as necessarily
representing the official policies or endorsements, either expressed or
implied, of the ODNI, IARPA, DARPA, or the U.S. Government. The U.S. Government
is authorized to reproduce and distribute reprints for Governmental
purposes notwithstanding any copyright annotation thereon.
\end{acknowledgements}

\bibliography{refs}

\appendix 
\section{Notes on divided differences} \label{app:divDiff}
We provide below a brief summary of the concept of divided differences which is a recursive division process. This method is typically encountered when calculating the coefficients in the interpolation polynomial in the Newton form.

The divided differences~\cite{dd:67,deboor:05} of a function $F(\cdot)$ is defined as
\beq\label{eq:divideddifference2}
F[x_0,\ldots,x_q] \equiv \sum_{j=0}^{q} \frac{F(x_j)}{\prod_{k \neq j}(x_j-x_k)}
\eeq
with respect to the list of real-valued input variables $[x_0,\ldots,x_q]$. The above expression is ill-defined if some of the inputs have repeated values, in which case one must resort to a limiting process. For instance, in the case where $x_0=x_1=\ldots=x_q=x$, the definition of divided differences reduces to: 
\beq
F[x_0,\ldots,x_q] = \frac{F^{(q)}(x)}{q!} \,,
\eeq 
where $F^{(n)}(\cdot)$ stands for the $n$-th derivative of $F(\cdot)$.
Divided differences can alternatively be defined via the recursion relations

\bea\label{eq:ddr}
&&F[x_i,\ldots,x_{i+j}] \\\nonumber
&=& \frac{F[x_{i+1},\ldots , x_{i+j}] - F[x_i,\ldots , x_{i+j-1}]}{x_{i+j}-x_i} \,,
\eea 
with $i\in\{0,\ldots,q-j\},\ j\in\{1,\ldots,q\}$ with the initial conditions
\beq\label{eq:divideddifference3}
F[x_i] = F(x_{i}), \qquad i \in \{ 0,\ldots,q \}  \quad \forall i \,.
\eeq
A function of divided differences can be defined in terms of its Taylor expansion. In the case where $F(x)=\e^{-\beta x}$, we have
\beq
\e^{-\beta [x_0,\ldots,x_q]} = \sum_{n=0}^{\infty} \frac{(-\beta)^n [x_0,\ldots,x_q]^n}{n!} \ . 
\eeq 

\section{Sign of $\e^{-\beta[E_0,\ldots, E_q]}$\label{app:posDivDiff}}

We note that $\e^{[x_0,\ldots, x_q]}$ is positive for any set of inputs $x_0,\ldots,x_q$~\cite{Farwig1985,GUPTA2020107385}. 
Setting $x_j=-\beta E_j$ for $j=0,\ldots,q$ yields $\e^{[-\beta E_0,\ldots, -\beta E_q]}>0$.  We next prove that
\beq
\e^{[-\beta E_0,\ldots, -\beta E_q]} = (-\beta)^q \e^{[E_0,\ldots, E_q]} \,.
\eeq
This immediately follows from the definition of divided differences. Explicitly:
\bea
&&\e^{[-\beta E_0,\ldots, -\beta E_q]}= \sum_j \frac{\e^{-\beta E_j}}{\prod_{k \neq j} (E_j- E_k)} \\\nonumber
&=& (-\beta)^q \sum_j \frac{\e^{-\beta E_j}}{\prod_{k \neq j} (-\beta E_j+\beta E_k)} =  (-\beta)^q \e^{[E_0,\ldots, E_q]} \,.
\eea
It follows then that 
\beq
\sgn \,\e^{[-\beta E_0,\ldots, -\beta E_q]}  = \sgn \, (-1)^q \,.
\eeq

\section{Positivity of $\cos m x$\label{app:cosmx}}

We show that the only $x \in [0,2\pi)$ for which \hbox{$\cos m x \geq 0$} for every natural number $m$ is $x=0$.
Plugging $x=0$, we obtain $\cos 0=1>0$ for every $m$. Next we show that for any $x \in (0,2\pi)$ there is an $m$ such that  $\cos m x <0$. 
We break down the statement to three cases: (i) If $x$ is in the interval $(\pi/2,3 \pi/2)$, then $\cos x$ is already negative. (ii) If $x \in (0,\pi/2]$ then consider the smallest $m$ for which $m x>\pi/2$ . We have $(m-1)x \leq \pi/2$ and $mx=(m-1)x+x>\pi/2$. Since  $(m-1)x \leq \pi/2$ and $0<x \leq \pi/2$ then $m x$ is necessarily in $(\pi/2, \pi]$ in which case $\cos mx<0$. (iii)  If $x \in [3\pi/2,2\pi)$ then consider the smallest $m$ for which  $(mx  \mod 2 \pi)< 3 \pi/2$. Since $((m-1)x \mod 2\pi) \geq 3 \pi /2$ and  $3\pi/2\leq x <2 \pi$ then $m x$ is necessarily in $[\pi, 3\pi/2)$ and so $\cos mx<0$.
\end{document}